\documentclass[final,5p,times,twocolumn,authoryear]{elsarticle}

\usepackage{amssymb}
\usepackage{lipsum}

\usepackage{amsmath}
\usepackage{hyperref}

\journal{High Energy Astrophysics}

\begin{document}

\begin{frontmatter}



\title{A doublet of cosmological models to challenge the $H_0$ tension in the Pantheon Supernovae Ia catalog}

\author[1,2]{B. De Simone\corref{cor1}}
\cortext[cor1]{Corresponding author, bdesimone@unisa.it}
\affiliation[1]{organization={Dipartimento di Fisica "E.R. Caianiello", Università di Salerno},
            addressline={Via Giovanni Paolo II, 132},
            city={Fisciano (SA)},
            postcode={84084},
            country={Italy}}

\affiliation[2]{organization={INFN Gruppo Collegato di Salerno - Sezione di Napoli. c/o Dipartimento di Fisica "E.R. Caianiello", Università di Salerno},
            addressline={Via Giovanni Paolo II, 132},
            city={Fisciano (SA)},
            postcode={84084},
            country={Italy}
            }

\author[3,9]{M. H. P. M. van Putten\corref{cor2}}
\cortext[cor2]{Corresponding author, mvp@sejong.ac.kr}
\affiliation[3]{organization={Department of Physics and Astronomy, Sejong University},
            addressline={98 Gunja-Dong, Gwangjin-gu, Seoul},
            city={Seoul},
            postcode={143-747},
            country={South Korea}}

\author[4,5,6,7,8]{M. G. Dainotti}
\affiliation[4]{organization={Division of Science, National Astronomical Observatory of Japan},
            addressline={2-21-1 Osawa, Mitaka},
            city={Tokyo},
            postcode={181-8588},
            country={Japan}}
\affiliation[9]{organization={INAF-OAS}, 
            addressline={Bologna via P. Gobetti 101 I-40129},
            city={Bologna},
            postcode={I-40129},
            country={Italy}}

\affiliation[5]{organization={The Graduate University for Advanced Studies (SOKENDAI)},
            addressline={Shonankokusaimura, Hayama, Miura District},
            city={Kanagawa},
            postcode={240-0115},
            country={Japan}}

\affiliation[6]{organization={Space Science Institute},
            addressline={4765 Walnut St Ste B},
            city={Boulder},
            postcode={80301},
            state={CO},
            country={USA}}            

\affiliation[7]{organization={Nevada Center for Astrophysics, University of Nevada},
            addressline={4505 Maryland Parkway},
            city={Las Vegas},
            postcode={89154},
            state={NV},
            country={USA}}

\affiliation[8]{organization={Bay Environmental Institute},
            addressline={P.O. Box 25},
            city={Moffett Field},
            postcode={94035},
            state={CA},
            country={USA}}

\author[1,2]{G. Lambiase}

\begin{abstract}
$\Lambda$CDM provides a leading framework in the interpretation of modern cosmology.
Nevertheless, the scientific community still struggles with many open problems in cosmology. Among the most noticeable ones, 
the tension in the Hubble constant $H_0$ is particularly intriguing, 
prompting a wide range of possible solutions. 
In the present work, the flat scale-free cosmology ($S$CDM) of Maeder (2017) is tested for $H_0$ tension in fits to the Pantheon sample of Supernovae Ia. The Pantheon sample is a collection of 1048 SNe Ia, which formally defines $H_0=H(0)$ by extrapolation to redshift zero of data over positive redshifts $z>0$. Here, we  consider $H_{0,k}$ in fits of $S$CDM over $k$ equally sized bins of sub-samples of mean redshift $z_k$. To quantify a trend in $H_{0,k}$ with $z_k$, the results are fit by $f(z)=H'_0/(1+z)^\alpha$ in the two parameters $H'_0$ and $\alpha$. This approach tests for model imperfections or data biases by trends that are inconsistent with zero. 
Our findings show a decreasing trend inconsistent with zero at $5.3 \sigma$ significance, significantly more so than in $\Lambda$CDM. 
These results are further confirmed in Pantheon+.
It appears that a solution to $H_0$ tension is to be found in models with a deceleration parameter $q_0<q_{0,\Lambda}$ below that of $\Lambda$CDM, rather than $q_{0,\Lambda} < q_{0,S}<0$ satisfied by $S$CDM.
\end{abstract}


\begin{keyword}
Cosmology \sep Supernovae Ia \sep Hubble tension \sep Hubble constant



\end{keyword}

\end{frontmatter}




\section{Introduction}
\label{introduction}
Modern cosmology is most widely explained by the flat $\Lambda$-Cold Dark Matter model, commonly known as $\Lambda$CDM. This framework includes Dark Energy (DE), driving cosmic expansion represented by the constant $\Lambda$; Cold Dark Matter (CDM), a non-relativistic component; and an assumption of spatial flatness. The $\Lambda$CDM model is broadly accepted among scientists due to its demonstrated ability to account for the accelerated expansion of the universe. This was initially established by observations of Type Ia Supernovae (SNe Ia), one of the most precise “standard candles” available \citep{Riess1998,Perlmutter1999,2014A&A...568A..22B,Scolnic2018P,2022ApJ...938..113S}. Standard candles are astrophysical objects with a known intrinsic luminosity or a strong luminosity-distance relationship, providing reliable distance measurements. Cepheid stars are another key standard candle used to calibrate nearby SNe Ia distances \citep{2019ApJ...886L..27R,2021ApJ...911...12J,Perivolaropoulos2021,2024arXiv240302801A,2023PhLB..84037886T,2022ApJ...939...89B,2022ApJ...940...64Y}. Using these probes, the SH0ES team estimates a local $H_0$ value of $73.04 \pm 1.04$ \citep{2022ApJ...934L...7R}.

While SNe Ia serve as powerful cosmological tools, they are observationally limited: current data, even with the James Webb Space Telescope, reach only up to redshift $z = 2.90$ \citep{2024ApJ...971L..32P}. Therefore, other cosmological probes are essential to bridge the gap between SNe Ia observations and the early universe.

Alternative “standardizable” candles, such as Gamma-Ray Bursts (GRBs, \citealt{10.1111/j.1365-2966.2009.15456.x,2017ApJ...848...88D,2022MNRAS.516.1386C,2022MNRAS.510.2928C,2022MNRAS.514.1828D,2023MNRAS.521.3909B,2023MNRAS.518.2201D,2024arXiv240810707B,Bargiacchi2024arXiv240810707B})
and quasars (QSOs, \citealt{2023ApJ...950...45D,2024PDU....4401428D})
, extend the Hubble diagram to much higher redshifts, with GRBs observed up to $z=9.4$ \citep{Cucchiara2011} and QSOs up to $z=10.1$ \citep{2024ApJ...960L...1N}. GRB-based $H_0$ measurements yield $H_0 = 73.225 \pm 3.307$ without evolutionary correction and $H_0 = 72.869 \pm 2.921$ with it \citep{2023MNRAS.518.2201D}. For QSOs, calibration with SNe Ia provides $H_0 = 73.76 \pm 2.18$ without and $69.82 \pm 2.27$ with correction for evolution \citep{2023ApJS..264...46L}.

Despite its success, the $\Lambda$CDM model faces unresolved issues. The most significant is the so-called $H_0$ tension—an approximately $5\sigma$ discrepancy between the local $H_0$ estimated with SNe Ia and Cepheids \citep{2022ApJ...934L...7R} and the cosmological $H_0$ derived from the Cosmic Microwave Background (CMB), which is $67.4 \pm 0.5$ \citep{2020A&A...641A...6P}.  
If $\Lambda$CDM is the correct model and in the absence of systematic errors in observations, then $H_0$ would be the same in the $\Lambda$CDM fit data in the Local Distance Ladder (LDL) and the {\em Planck}-$\Lambda$CDM analysis of the CMB.

Motivated by the $H_0$ tension between LDL and {\em Planck}, researchers have explored various theoretical approaches. Proposals include alternative gravity theories, such as teleparallel gravity \citep{Najera2021,Ren2021} and modified gravity models \citep{2021PhRvD.104j3534A,Ambjorn2021,2022ApJ...935..156B,2022EPJC...82..418G,2023PDU....3901153B,2024PDU....4601652E,2024PDU....4301407S}. Changes to standard DE models \citep{Colgain2021DDE,2022PhRvD.106j3522G,2024arXiv241107060M} and alternative DM models \citep{Hansen2021,2023JCAP...11..005B,2024arXiv240303484K,2024FrASS..1113816N} are among the most prominent, together with models of DM-DE interaction \citep{2024arXiv240415977M}. Other solutions involve dark radiation, interactive radiation models \citep{Ghosh2021}, early- and late-time cosmology modifications \citep{2021arXiv210510425V,Ye2021a,2022JCAP...04..042A}, and quantum cosmologies \citep{Novikov2016fzd,2022PhRvD.106f3520A,2023PhRvD.108i5033C,2024EPJC...84..121K,2024arXiv240310865V}. Proposals also include local inhomogeneities \citep{Fanizza2021,2021arXiv210503003T,2024JCAP...05..126M}, exotic particles \citep{Fung2021,Cuesta2021}, fundamental physics variations \citep{Sola2021}, and modifications to the gravitational constant \citep{Marra2021,alestas2022}.
The cosmographic approach, which is scale factor-independent, is an interesting alternative tool for the study of the cosmos and can give more insights into the cosmological parameter tensions \citep{2022PhRvD.106l3523P,2024SSRv..220...48B}.
Of particular interest are the proposals where the role of the other cosmological parameters is investigated to assess how the Hubble tension can be reduced \citep{2020PhRvD.102b3518V,2024arXiv240804530P}.
A more exhaustive review of the proposals aimed to tackle the $H_0$-tension is reported in the following works:
\citet{2021arXiv210301183D,2021arXiv210505208P,Perivolaropoulos2021review,Abdalla2022JHEAp..34...49A,Moresco2022,2023Univ....9..501C,2023PDU....4201348P,2023Univ....9..393V,2024PhRvD.109l3545G,2024arXiv240811031P,2024ARA&A..62..287V}.

Possible solutions to an apparent evolution of $H_0$ with the redshift scale may be the removal of astrophysical biases or redshift evolution effects, if present, or alternative cosmologies \citep{2020PhRvD.102j3525K,Dainotti2021hubble,Krishnan2021a,Dainotti2022hubble,10.1093/pasj/psac057,2022MNRAS.517..576H,2023A&A...674A..45J,DainottiBargiacchi2022c,
2022PhRvD.106f3520A,2023PhRvD.108i5033C,2024EPJC...84..121K,2024arXiv240310865V}. In the present work, the approach of \citet{Dainotti2021hubble} is considered to investigate a flat scale-invariant cosmology \citep{Maeder2017,Jesus2017} as an alternative to $\Lambda$CDM. 

The authors in \citet{Dainotti2021hubble} propose a novel analysis of late-time cosmological models using the Pantheon sample \citep{Scolnic2018P}. This analysis considers $H_{0,k}$ estimates extracted from model fits over bins with running mean redshift $z_k$. Trends in $H_{0,k}$ inconsistent with zero would indicate a model-imperfection (failing the null-hypothesis). To this end, the Pantheon sample is split into equally populated bins, here by partitions into a total of $K$ bins, $K$ =3, 4, 20, and 40. 
In each bin, Bayesian analyses are performed on $\Lambda$CDM and $S$CDM through emulated by $w_{0}w_{a}$CDM \citep{CPL1,CPL2}. To identify a trend, the resulting $H_{0,k}$ ($k=1,2,... K$) are fit with $f(z)=H'_0/(1+z)^\alpha$, parameterized by $H'_0$ and $\alpha$ - considering that the confidence level in a possible trend is quantified by $\alpha/\sigma_{\alpha}$.

A trend may be attributed to a redshift drift in systematic errors in supernova data - the SNe Ia stretch parameter \citet{Nicolas2021}. If not due to systematics, a trend points to the need for an alternative cosmological model, such as
$f(R)$ modified gravity theories \citep[e.g.][]{2011cqvz.book..227C,Dainotti2021hubble,Dainotti2022hubble,Schiavone2023,MONTANI2024101486,2024arXiv240801410S}.
To reflect on the $S$CDM analysis, the results of this test of $\Lambda$CDM obtained previously \citep{Dainotti2021hubble} are included below. 

The structure of the present manuscript follows. In Section \ref{sec:cosmology}, $\Lambda$CDM and $S$CDM are briefly described. 
In Section \ref{sec:data}, the Pantheon sample of SNe Ia data are introduced with relevant observables for cosmological analysis.
In Section \ref{sec:method}, a description of binning is provided.
The results of this analysis in Section \ref{sec:results} are presented on the conclusions and interpretation in Section \ref{sec:conclusions}.

In what follows, the Hubble constant $H_0$ is expressed in 
units of km\,s$^{-1}$Mpc$^{-1}$.

\section{The standard $\Lambda$CDM cosmology}
\label{sec:cosmology}

$\Lambda$CDM embodies the Copernicus principle of a homogeneous and isotropic universe, expressed by the Friedmann equations \citep{Weinberg2008} in terms of an evolving scale factor $R=R(t)$ of otherwise flat Minkowski spacetime.
Central to studies of the Hubble expansion based on supernova data is the luminosity distance:

\begin{equation}
    d_L(z)=\frac{c(z+1)}{H_0}\int^{z}_{0}\frac{dz'}{\sqrt{\Omega_{M,0} (1+z')^3 + \Omega_{\Lambda,0}}},
    \label{eq:lumdist}
\end{equation}
parameterized by the dimensionless density parameters $\Omega_{i,0}$ at the present epoch (redshift zero) for total matter ($M$) and DE ($\Lambda$), neglecting radiation and assuming vanishing curvature. The distance moduli satisfy:

\begin{equation}
    \mu_{th}(z)=5\log_{10}d_L(z) + 25.
    \label{eq:distancemoduli}
\end{equation}

\subsection{The exact solution for the flat scale-invariant cosmology}\label{sec:SCDM}

The work of \citet{Maeder2017} suggests that the cosmic accelerating expansion phase can be explained through scale-invariance of empty space. This replaces $\Lambda$ with a variable contribution that involves the scale factor with no assumptions on DE 
\citep{Jesus2017}. 
The model is here labeled as $S$CDM. 
In this framework, the scale-invariant evolution of a three-flat universe satisfies:

\begin{equation}
    \dot{R}^2 Rt -2\dot{R}R^2 - Ct^2 = 0,
    \label{eq:scaleinvariant1}
\end{equation}
where the cosmic time $t$ is written in units of $t_0=1$ and $R_0=1$. Considering (\ref{eq:scaleinvariant1}) today, and
recalling $H_0 \equiv \dot{R}/R$ at $t_0=1$, 
the relation leads to:
\begin{equation}
    H^2_0-2H_0=C.
    \label{eq:H0flatscale}
\end{equation}

Here, only the positive solution can be considered in $H_0=1 \pm \sqrt{1+C}$ in keeping with the expanding universe. 
The scale factor satisfies
\begin{equation}
    R=tv=\biggr(\frac{(2H_0+C)t^3-C}{2H_0} \biggr)^\frac{2}{3},
    \label{eq:scaleinvariant2}
\end{equation}
In terms of $\Omega_{M,0}$, it becomes
\begin{equation}
    R=\biggr(\frac{t^3-\Omega_{M,0}}{1-\Omega_{M,0}}\biggr)^\frac{2}{3}.
    \label{eq:scaleinvariant3}
\end{equation}
To compute the exact solution of the scale-invariant cosmology the $H(t)$ is called into question. Using the Eq. \ref{eq:scaleinvariant3}:
\begin{equation}
    H(t) \equiv \dot{R}/R= \frac{2t^2}{t^3-\Omega_{M,0}}.
    \label{eq:Hubbleparameter}
\end{equation}
Solving (\ref{eq:scaleinvariant3}) in terms of $t$ and substitution in (\ref{eq:Hubbleparameter}), 
the scale-invariant Hubble parameter returns the following luminosity distance:
\begin{equation}
    d_L=\frac{c(z+1)}{H_0}\int^{z}_{0}\frac{dz'}{[\Omega_{M,0}(1+z')^\frac{9}{4}+(1-\Omega_{M,0})(1+z')^\frac{3}{4}]^\frac{2}{3}}.
    \label{eq:jesuslumdist}
\end{equation}
From (\ref{eq:distancemoduli}-\ref{eq:jesuslumdist}), it is readily seen that $H_0$ is crucial in the estimation of the cosmological observables. This is accompanied by the \emph{deceleration parameter} $q=-1+\left(1+z\right)H^\prime(z)/H(z)$, providing a dimensionless expression of the slope of the Hubble diagram at redshift $z)$. At redshift zero, it reduces to
\begin{equation}
    q_0=-1+\frac{H^\prime_0}{H_0}.
    \label{eq:decelerationparam}
\end{equation}
For $\Lambda$CDM and $S$CDM, we have, respectively, 
\begin{eqnarray}
 q_0^\Lambda=\frac{1}{2}\Omega_{M,0}-\Omega_{\Lambda,0} \simeq -\frac{1}{2},\,\,\,
 q_0^S = \Omega_{M,0} - \frac{1}{2} \simeq -\frac{1}{6}
\label{EQN_q0}
\end{eqnarray}
for a canonical values $\Omega_{M,0}\simeq 1/3$. Crucially, (\ref{EQN_q0}) shows
\begin{equation}
    q_0^\Lambda < q_0^S < 0.
    \label{EQN_q0b}
\end{equation}

\begin{figure}
    \centering
    \includegraphics[width=1.1\linewidth]{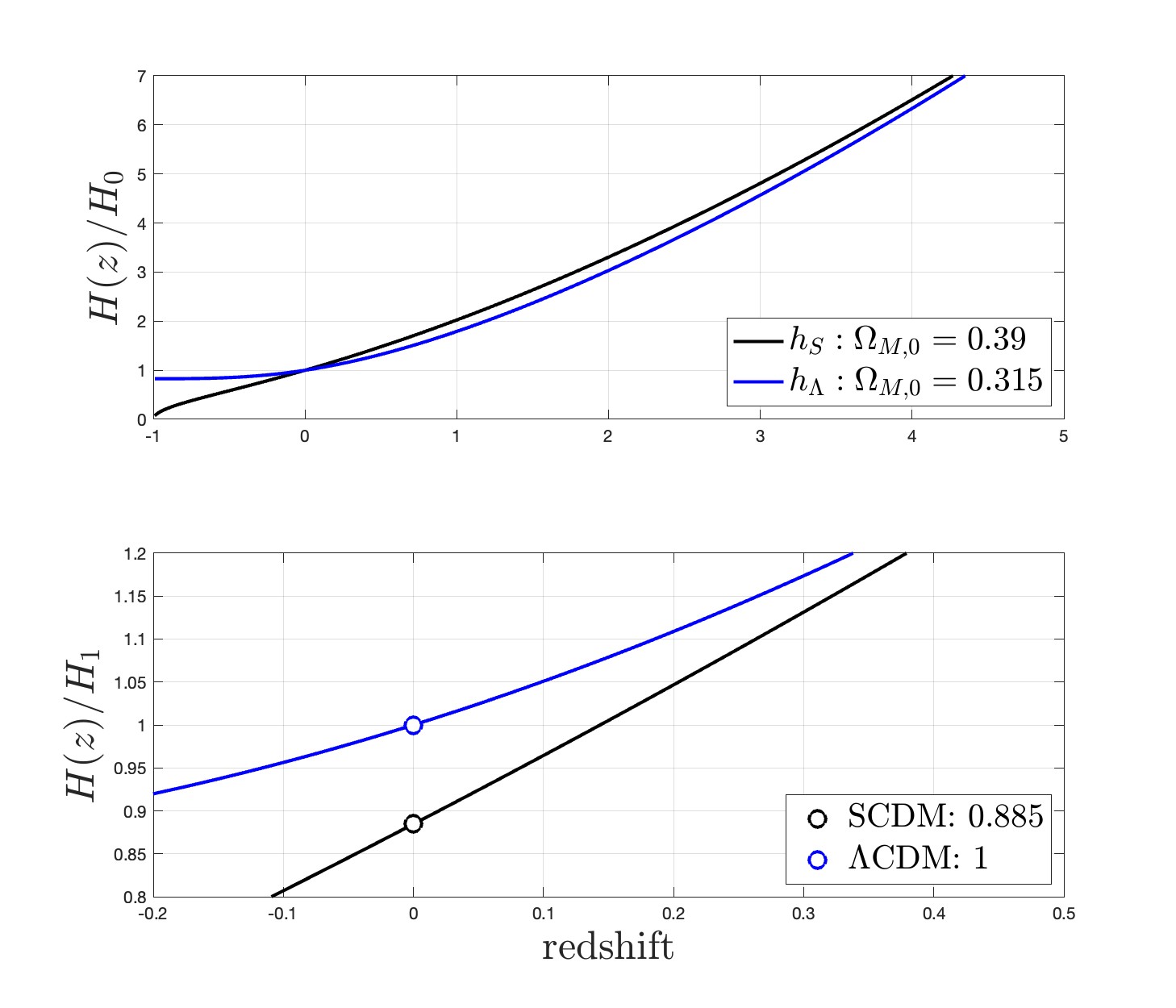}
    \caption{The $H(z)/H_0$ plot for the $\Lambda$CDM and $S$CDM (upper panel, blue and black line, respectively) and the $H(z)/H_1$ plot for the same two models (lower panel, blue and black line, respectively). The values of $\Omega_{M,0}$ are $0.315,0.39$ for the $\Lambda$CDM and $S$CDM, respectively (plot after \citealt{PhysRevD.104.083511}).}
    \label{fig:H(z)/H0}
\end{figure}

In Fig. \ref{fig:H(z)/H0}, the upper panel contains the $\Lambda$CDM Hubble parameter divided by $H_0$ in blue, while the ratio between the same quantity and the $S$CDM Hubble parameter is reported in black. In the lower panel, $H(z)/H_1$ is reported for the same models, given that $H_1 \equiv H(z=1)$: due to the nature of $H_1$, it is obvious that $H(z)/H_1 =1$ at $z=1$ for the $\Lambda$CDM, while in the case of $S$CDM, the ratio assumes the value $\sim 0.885$. The two panels share the same meaning of colors.
In Fig. \ref{fig:H(z)/H0}, the values of $\Omega_{M,0}$ are fixed to $0.315$ and $0.39$ for the $\Lambda$CDM and $S$CDM, respectively.

\section{The Pantheon and Pantheon+ supernova samples} \label{sec:data}

SNe Ia provide excellent data to test the various cosmological models, notably so $\Lambda$CDM. 
The advantage of these astrophysical probes for testing the Hubble expansion is their almost uniform peak magnitude, given the precise threshold of critical mass derived by the Chandrasekhar limit. 
SNe Ia are detected up to redshift $z=2.26$ in previous years but, considering the recent observations from the James Webb Space Telescope, SNe Ia reaches $z=2.90$ \citep{2024ApJ...971L..32P}. 
In the data release of \citet{Scolnic2018P}, SNe Ia are reported together with their light-curve parameters and observed distance moduli ($\mu_{obs}$) with uncertainties (including both systematic and statistical contributions). 
The $\mu_{obs}$ value is computed exploiting a modification of the Tripp formula \citep{Tripp1998}. The SNe Ia light-curve parameters are fixed and the fiducial absolute magnitude ($M$) is re-calibrated according to the local $H_0$, as in \citet{Dainotti2021hubble}. 
In this way, the tabulated values of $\mu_{obs}$ are considered for the cosmological analysis.
The statistical uncertainties on the $\mu_{obs}$ are included in the matrix $D_{stat}$, while the systematic ones are in $C_{sys}$. The total covariance matrix, that is used in the binning approach, is obtained by their sum:
\begin{equation}
    \mathcal{C}=D_{stat}+C_{sys}.
    \label{eq:covmatrix}
\end{equation}
The same structure of the data release can be found in the Pantheon+ catalog \citep{Scolnic2022PP}.

\section{Methodology: binned trend analysis} \label{sec:method}

The methodology applied in this work follows \citet{Dainotti2021hubble}. The Pantheon sample of 1048 SNe Ia is partitioned into $K$ equally sized bins, choosing $K=3$,$4$,$20$, and $40$. This binning method proved itself to be effective to test for consistency in cosmological parameters estimation. In \citet{2024PDU....4401428D,galaxies12010004}, the selection of QSO bins size is performed such that the statistics in each of the $k=1,2,... K$ bins is not reduced significantly and the local features of the cosmological parameters in each bin are still observable, not dominated by scatter (statistical fluctuations). 

It could be argued that bin splitting may induce some boundary effect at the edges of the redshift-ordered bins. 
Considering a more coarse-grained division of the SNe Ia sample, say $K=4$ bins, any SN Ia that appears to be an outlier for the Hubble diagram would be included in a relatively large set of SNe, whose effect may be smeared out that otherwise might affect estimates over finer bins divisions (larger $K$). As observed in \citet{Dainotti2021hubble}, and further below in the present analysis, the parameters of interest that are estimated from the fitting results obtained over $K$ bins appear independent of course graining (defined by $K$) in the present binning of the Pantheon data, essentially within $1 \sigma$.
In the work by Dainotti et al. (in preparation), furthermore, binning by equal volume instead also shows no change in trends in analyzing the Pantheon sample. 
To create the SNe Ia bins, an \textit{ad hoc} code is written in \emph{Wolfram Mathematica}. The code extracts sequentially the SNe starting from a given position (according to the redshift) and proceeds with a chosen number of SNe where each bin is equally populated. With our code, it is possible to extract, for example, for 4 redshift bins in the Pantheon sample, $1048/4=262$ SNe Ia in each bin with an increasing redshift order. For each redshift bin, the elements of the matrix for each SNe Ia are moved in the correct position accordingly, namely the code extracts the corresponding submatrices, including also systematic errors. The extracted full covariance matrix contains elements associated with the SNe Ia with the redshifts within the considered bin.
This code property can be beneficial since the aim is to extract redshift-ordered bins, providing the observed distance moduli, the redshift, and the elements of the inverse covariance matrix. The code will be available for the readers upon reasonable request.
The $M$ parameter requires a calibration to the local values of the $H_0$. 
To this end, adopting the same approach as \citet{Dainotti2021hubble}, $M$ is estimated in the first bin $(k=1$) of each division is normalized to the same value:

\begin{eqnarray}
H_{0,1} = 73.5 \,{\rm km\,s}^{-1}{\rm Mpc}^{-1},
\label{EQN_H01}
\end{eqnarray}
compatible with the majority of local constraints.
Estimates of $M$ are then kept fixed for the remaining bins in each partition of size $K$. 
The values of $M$ for $\Lambda$CDM used here are reproduced from \citet{Dainotti2021hubble}, included in Table \ref{tab:H0fits} alongside the new results for $S$CDM. After fixing $M$, for each bin $k=1,2,... K$, the sub-covariance matrix and the values of $\mu_{obs}$ are extracted. 
With these elements, the $\chi^2$ test is be performed with 
\begin{equation}
    \chi^2=\Delta \mu^{T} \mathcal{C}^{-1} \Delta \mu,
    \label{eq:chi2normal}
\end{equation}
where $\Delta \mu = \mu_{obs}-\mu_{th}$, where $\mu_{th}$ is the theoretical distance moduli with a given cosmological setting. 
Through the Markov Chain Monte Carlo (MCMC) approach the posterior distributions for $H_0$ are sampled together with their mean values and standard deviations, assuming a large range of flat priors such that $60 \leq H_0 \leq 80$. The total matter density is fixed to $\Omega_{M,0}=0.298$ according to the value in \citet{Scolnic2018P} since we use this data release.
The mean and the standard deviation of the $H_{0,k}$ resulting from each bin $k$ are then given a weighted non-linear fit 
\begin{equation}
    f(z)=\frac{H'_0}{(1+z)^\alpha}
    \label{eq:h0z}
\end{equation}
in the two parameters $H'_0$ and $\alpha$. The index $\alpha$ is our diagnostic for a possible trend in $H_{0,k}$, providing a test of the null-hypothesis of no trend for a correct cosmological model, based on 
$H_{0,k}$ over each bin with mean $z_k$ $(k=1,2,... K)$ including the the $1 \sigma$ STD in the estimate $H_{0,k}$.
After the fit (\ref{eq:h0z}), a possible trend is quantified also by the percentage difference between $H_{0,k}$ and $H_{0,K}$, 
\begin{equation}
    \Delta_\% = \frac{|H_{0,K}-H_{0,1}|}{\sqrt{\sigma^2_{H_{0,K}}+\sigma^2_{H_{0,1}}}},
    \label{eq:percentagediff}
\end{equation}
where the STDs refer to the uncertainties in the model fits to each individual bin. 

\section{Discussion of the results} \label{sec:results}

\subsection{Results for the Pantheon sample}\label{sec:resultspantheon}

Table \ref{tab:H0fits} summarizes the current results for $S$CDM and the previous $\Lambda$CDM side-by-side based on the fits (\ref{eq:h0z}). 
The latter is included from \citet{Dainotti2021hubble}, whose trend measured by $\alpha$ is consistent with zero (the null-hypothesis) within $2.0 \sigma$. 
In contrast, $S$CDM shows a trend by $\alpha$ inconsistent with zero at a significance of $5.3 \sigma$ (3 bins) with higher significance inferred from analysis with a larger number of bins. For $S$CDM, $\alpha$ estimates for different $K$ are consistent withing $1 \sigma$.

For $\Lambda$CDM, the percentage differences $\Delta_\%$ between $H_{0,1}$ and $H_{0,K}$ are modest, ranging from  $0.9 \sigma$ to $1.0 \sigma$. In contrast, $S$CDM shows a significant change at $4.4 \sigma$ to $10.9 \sigma$ significance, which result shows no appreciable dependence on the number of bins $K$.

This outcome for significant trending and large $\Delta_\%$ in $S$CDM is expected based on Fig. \ref{fig:H(z)/H0} and (\ref{EQN_q0b}).
Recall that $H_{0,k}$ represents an extrapolation of data to the left ($z=0$) away from the mean $z_k>0$ of bin $k$. For the steep slope along an essentially straight-line Hubble diagram of $S$CDM, this extrapolation explains a drop that varies essentially linearly linearly with $z_k$, evident in Figs. \ref{fig:H0fit3binsP}, \ref{fig:H0fit4binsP}, \ref{fig:H0fit20binsP}, and \ref{fig:H0fit40binsP}.

{In contrast, the slope of the Hubble diagram of $\Lambda$CDM is relatively modest, based on Fig. \ref{fig:H(z)/H0} and (\ref{EQN_q0b}). 
This leaves a modest trending and relatively small $\Delta_\%$. While not statistically significant, these two quantities appear with a modest deviation from zero.

In Figs. \ref{fig:H0fit3binsP}, \ref{fig:H0fit4binsP}, \ref{fig:H0fit20binsP}, and \ref{fig:H0fit40binsP}, note that the fits to $\Lambda$CDM and $S$CDM intersect at $z=0$ at $H_0=73.5$ by construction, based on the normalization (\ref{EQN_H01}). For $K=3,4$, in fact, $\Delta_\%\simeq 10$ agrees with the drop of about 11\% anticipated by Fig. \ref{fig:H(z)/H0}. For more refined binning ($K=20$, 40), $\Delta_\%$ is smaller by a factor of about two, yet remains significant. 

In searches to address the $H_0$-tension problem by models beyond $\Lambda$CDM, it can be concluded, based on Table \ref{tab:H0fits} and (\ref{EQN_q0b}), that a preferred search is for models which $q_0<q_0^\Lambda$ rather than $q_0>q_0^\Lambda$ demonstrated by $S$CDM.

\begin{table}
\scriptsize
\begin{centering}
\begin{tabular}{|c|c|c|c|c|c|}
\hline
\multicolumn{6}{|c|}{{\bf Pantheon, 3 bins, Varying $H_0$}}\tabularnewline
\hline
Model & $H'_0$ & $\alpha$ & $\alpha/\sigma_{\alpha}$ & $\Delta_\%$ & $M \pm \sigma_M$ \\
& & & & &
\tabularnewline
\hline
$\Lambda$CDM & $73.577\pm0.106$ & $0.009\pm0.004$ & $2.0$ & $0.9$ & $-19.245\pm0.006$ 
\tabularnewline
\hline
$S$CDM & $73.939\pm0.359$ & $0.098\pm0.018$ & $5.3$ & $9.6$ & $-19.212\pm0.005$ \tabularnewline
\hline
\multicolumn{6}{|c|}{{\bf Pantheon, 4 bins, Varying $H_0$}}\tabularnewline
\hline
Model & $H'_0$ & $\alpha$ & $\alpha/\sigma_{\alpha}$ & $\Delta_\%$ & $M \pm \sigma_M$ \\
& & & & &
\tabularnewline
\hline
$\Lambda$CDM & $73.493\pm0.144$ & $0.008\pm0.006$ & $1.5$ & $0.9$ & $-19.246\pm0.008$ \tabularnewline
\hline
$S$CDM & $73.515\pm0.278$ & $0.096\pm0.011$ & $8.5$ & $10.9$ & $-19.291\pm0.007$ \tabularnewline
\hline
\multicolumn{6}{|c|}{{\bf Pantheon, 20 bins, Varying $H_0$}}\tabularnewline
\hline
Model & $H'_0$ & $\alpha$ & $\alpha/\sigma_{\alpha}$ & $\Delta_\%$ & $M \pm \sigma_M$ \\
& & & & &
\tabularnewline
\hline
$\Lambda$CDM & $73.222\pm0.262$ & $0.014\pm0.010$ & $1.3$ & $1.0$ & $-19.262\pm0.014$ \tabularnewline
\hline
$S$CDM & $72.580\pm0.180$ & $0.113\pm0.009$ & $12.3$ & $6.0$ & $-19.258\pm0.018$ \tabularnewline
\hline
\multicolumn{6}{|c|}{{\bf Pantheon, 40 bins, Varying $H_0$}}\tabularnewline
\hline
Model& $H'_0$ & $\alpha$ & $\alpha/\sigma_{\alpha}$ & $\Delta_\%$ & $M \pm \sigma_M$ \\
& & & & &
\tabularnewline
\hline
$\Lambda$CDM & $73.669\pm0.223$ & $0.016\pm0.009$ & $1.8$ & $0.7$ & $-19.250\pm0.021$ \tabularnewline
\hline
$S$CDM & $73.471\pm0.180$ & $0.119\pm0.009$ & $13.6$ & $4.4$ & $-19.234\pm0.023$ \tabularnewline
\hline
\end{tabular}
\caption{Table with the $f(z)$ fitting results for $\Lambda$CDM and $S$CDM in the Pantheon sample. First column: the cosmological model. Second column: the value of $H'_0 \pm \sigma_{H'_0}$ from the fitting. Third and fourth columns: the value of $\alpha \pm \sigma_{\alpha}$ and the number of $\sigma$ in which it is compatible with zero, respectively. Fifth column: the percentage difference in each case. Last column: the value of $M$ estimated from the first bin of each division. The values related to $\Lambda$CDM are reproduced from \citet{Dainotti2021hubble}.}
\label{tab:H0fits}
\par\end{centering}
\end{table}

\begin{figure}
    \centering
    \includegraphics[width=0.8\linewidth]{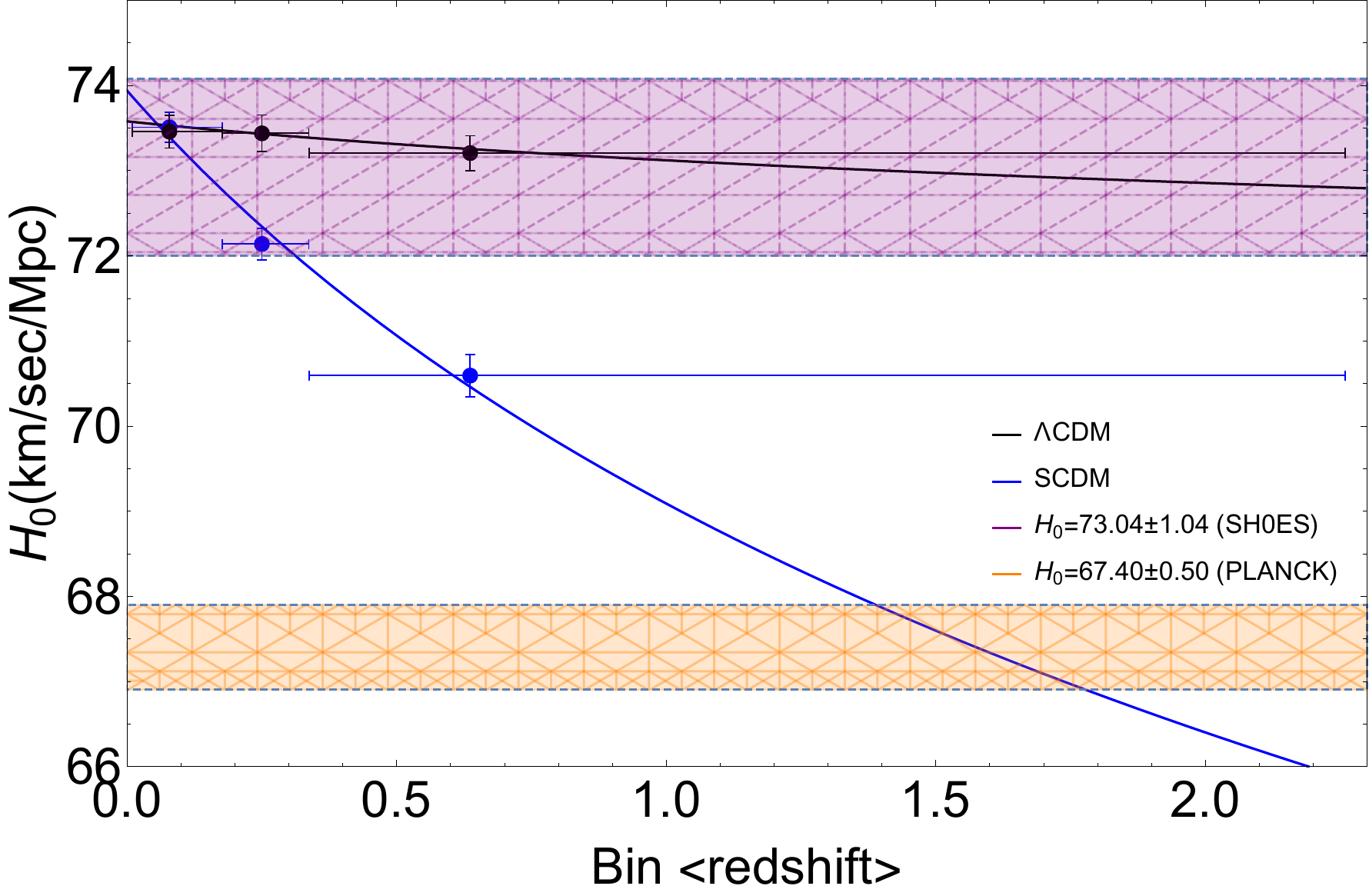}
    \caption{Fitting of $f(z)$ in the 3 bins cases: $\Lambda$CDM (black line and markers) and $S$CDM (blue line and markers). In purple, the confidence interval for $H_0$ estimated through SH0ES \citep{2022ApJ...934L...7R} is depicted, while in orange the same plot is performed considering $H_0$ from Planck CMB analysis \citep{2020A&A...641A...6P}. The $\Lambda$CDM data and fitting line are reproduced from \citet{Dainotti2021hubble}.}
    \label{fig:H0fit3binsP}
\end{figure}

\begin{figure}
    \centering
    \includegraphics[width=0.8\linewidth]{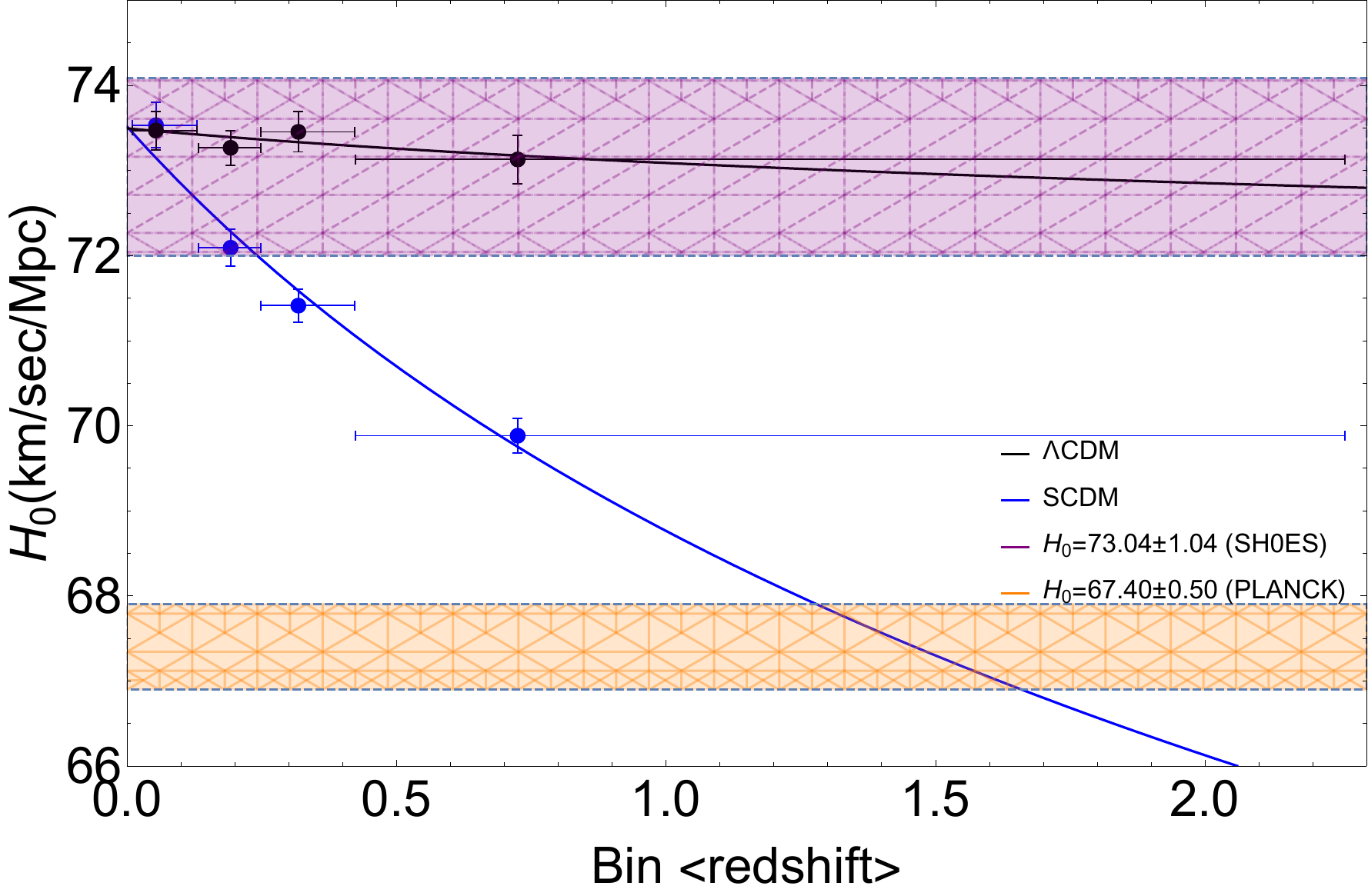}
    \caption{Fitting of $f(z)$ in the 4 bins cases: $\Lambda$CDM (black line and markers) and $S$CDM (blue line and markers). In purple, the confidence interval for $H_0$ estimated through SH0ES \citep{2022ApJ...934L...7R} is depicted, while in orange the same plot is performed considering $H_0$ from Planck CMB analysis \citep{2020A&A...641A...6P}. The $\Lambda$CDM data and fitting line are reproduced from \citet{Dainotti2021hubble}.}
    \label{fig:H0fit4binsP}
\end{figure}

\begin{figure}
    \centering
    \includegraphics[width=0.8\linewidth]{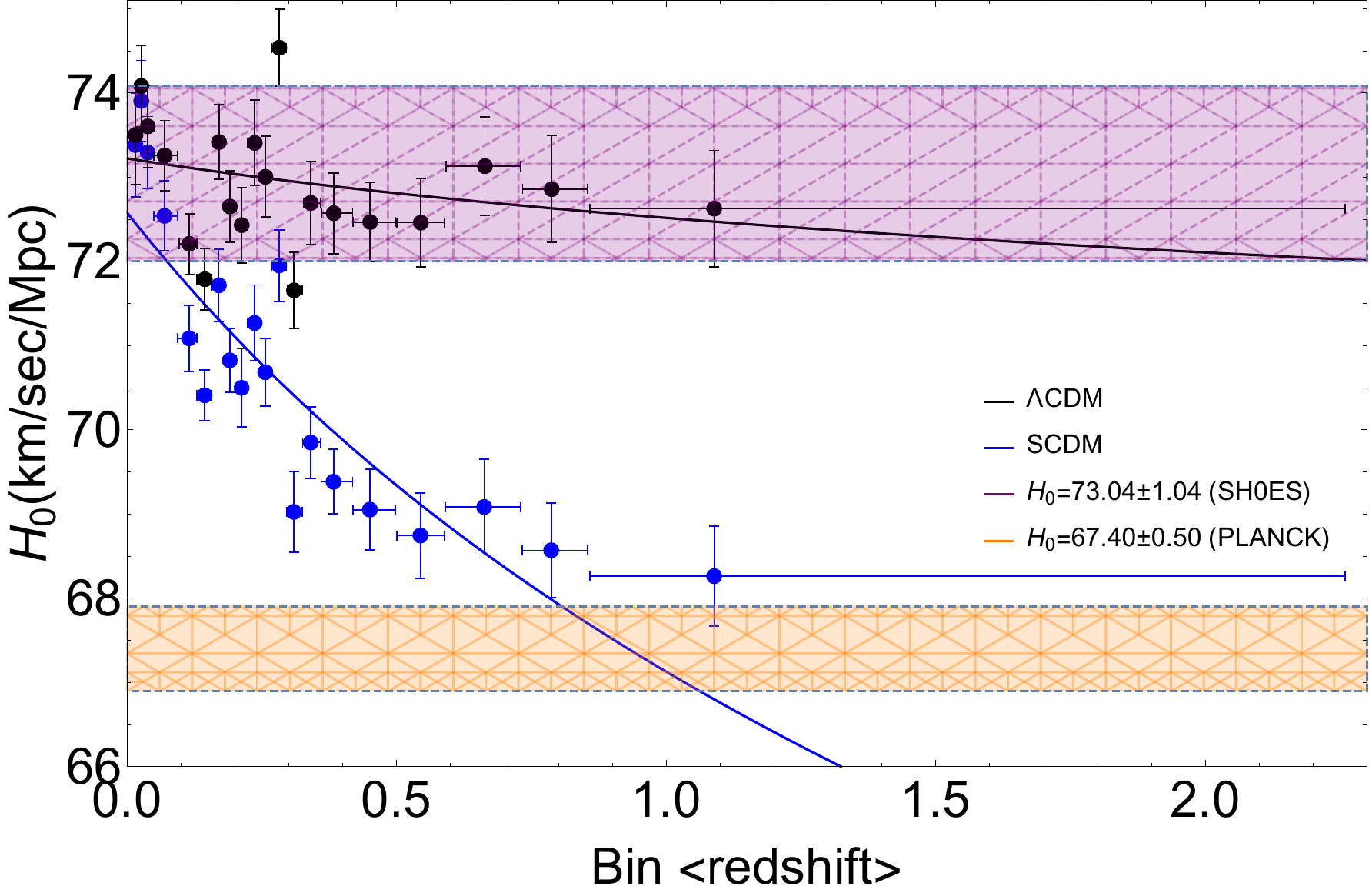}
    \caption{Fitting of $f(z)$ in the 20 bins cases: $\Lambda$CDM (black line and markers) and $S$CDM (blue line and markers). In purple, the confidence interval for $H_0$ estimated through SH0ES \citep{2022ApJ...934L...7R} is depicted, while in orange the same plot is performed considering $H_0$ from Planck CMB analysis \citep{2020A&A...641A...6P}. The $\Lambda$CDM data and fitting line are reproduced from \citet{Dainotti2021hubble}.}
    \label{fig:H0fit20binsP}
\end{figure}

\begin{figure}
    \centering
    \includegraphics[width=0.8\linewidth]{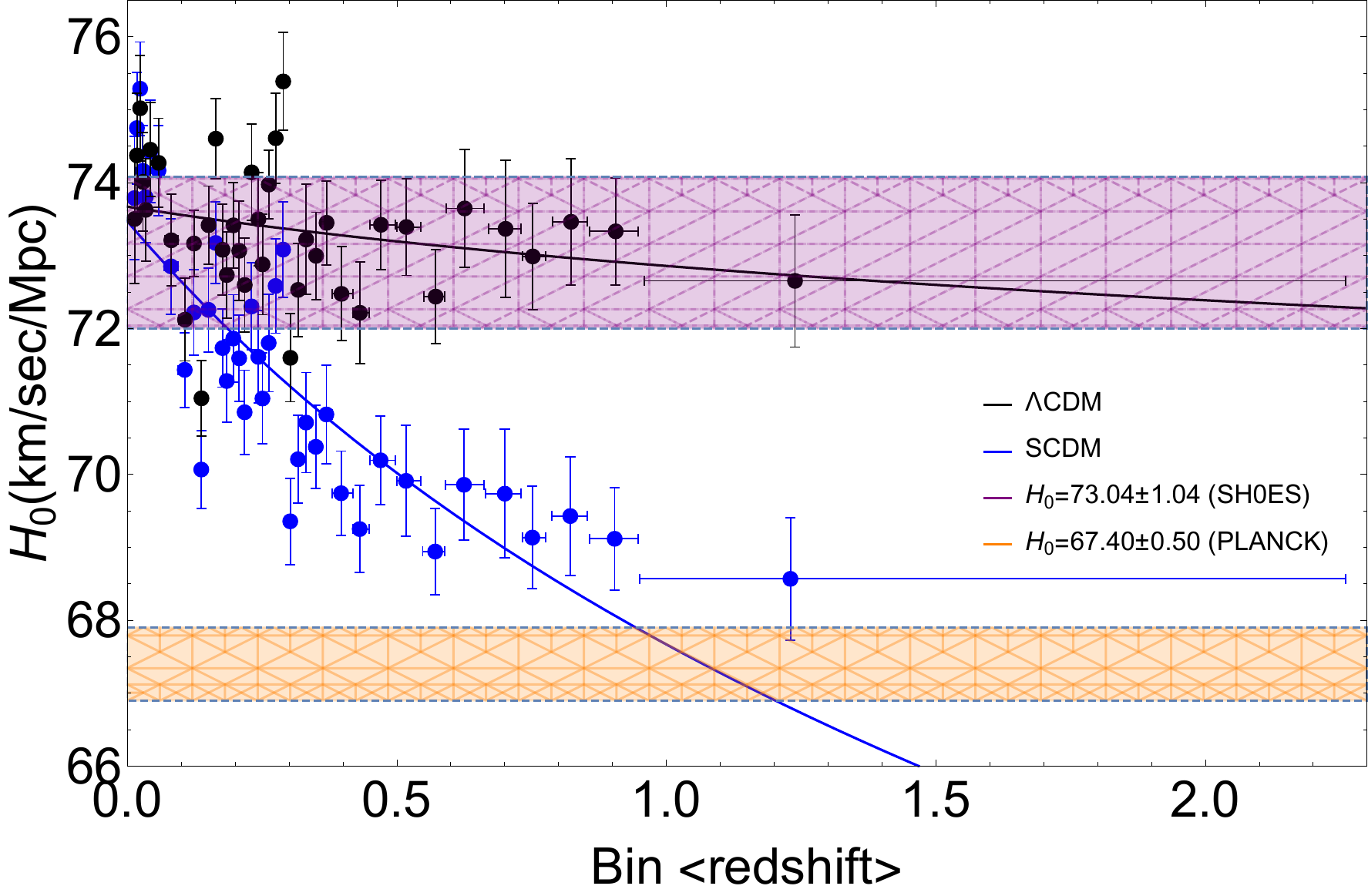}
    \caption{Fitting of $f(z)$ in the 40 bins cases: $\Lambda$CDM (black line and markers) and $S$CDM (blue line and markers). In purple, the confidence interval for $H_0$ estimated through SH0ES \citep{2022ApJ...934L...7R} is depicted, while in orange the same plot is performed considering $H_0$ from Planck CMB analysis \citep{2020A&A...641A...6P}. The $\Lambda$CDM data and fitting line are reproduced from \citet{Dainotti2021hubble}.}
    \label{fig:H0fit40binsP}
\end{figure}

\subsection{{\bf The $S$CDM model with Pantheon+}}\label{sec:resultspantheon+}
Here, the results of $S$CDM cosmology through Pantheon+ are presented. The Pantheon+ \citep{Scolnic2022PP} collects 1701 lightcurves (gathered from 1550 SNe Ia with spectroscopic confirmation) and expands the Pantheon sample. The redshift spans in $0.001 < z < 2.259$, being built up from 18 different surveys. The total matter density today in the Pantheon+ has been fixed to $\Omega_{M,0}=0.334$ following \citet{Brout2022}. As a remark, in the Pantheon+ analysis, the value of $M$ has not been estimated given that the values of $\mu_{obs}$ provided in the data release of \citet{Scolnic2022PP} have been computed using SH0ES 2022 Cepheid distances as local constraints \citep{2022ApJ...934L...7R}.
The plots in Figure \ref{fig:H0fitallbinsPP} are related to the 3, 4, 20, and 40 bins cases. According to the results, the $\alpha$ parameter that marks out any evolution of $H_0$ with redshift shows compatibility with zero in a range from $4.8 \sigma$ to $9.7 \sigma$. 
The results of these plots are reported in Table \ref{tab:H0fitsPantheon+}. Interestingly, all the $\alpha$ values are compatible in $1 \sigma$. Furthermore, considering the percentage differences between the local values of $H_0$ and the one at the farthest bin, the $\Delta_\%$ ranges from $1.8$ up to $7.0$. 
Looking at the $\Delta_\%$, it shows a decreasing trend with the number of bins. Interestingly, this is observed for the Pantheon+ case and not for the Pantheon one. Finally, the $\alpha$ values in the case of 20 and 40 $S$CDM bins with Pantheon+ provide a slower evolution (smaller $\alpha$ value) without the compatibility in $1 \sigma$ if compared to the same case of Pantheon sample. This could be due to the different calibration processes and data coverage behind the two SNe Ia catalogs. The report in Table \ref{tab:H0fitsPantheon+} suggests the same conclusion as the one drawn from the Pantheon sample inspection.


\begin{figure}
    \centering
    \includegraphics[width=0.75\linewidth]{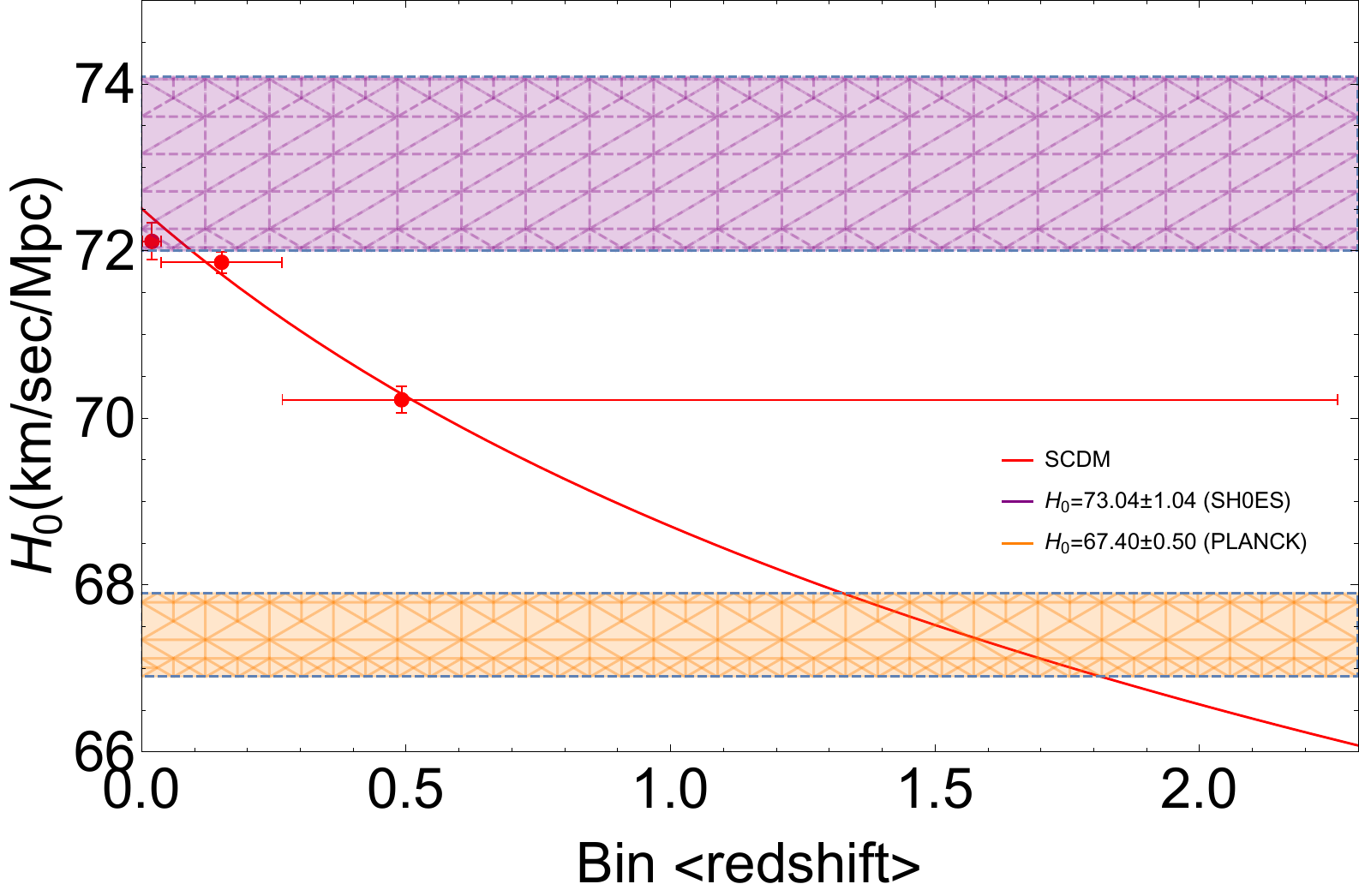}
    \includegraphics[width=0.75\linewidth]{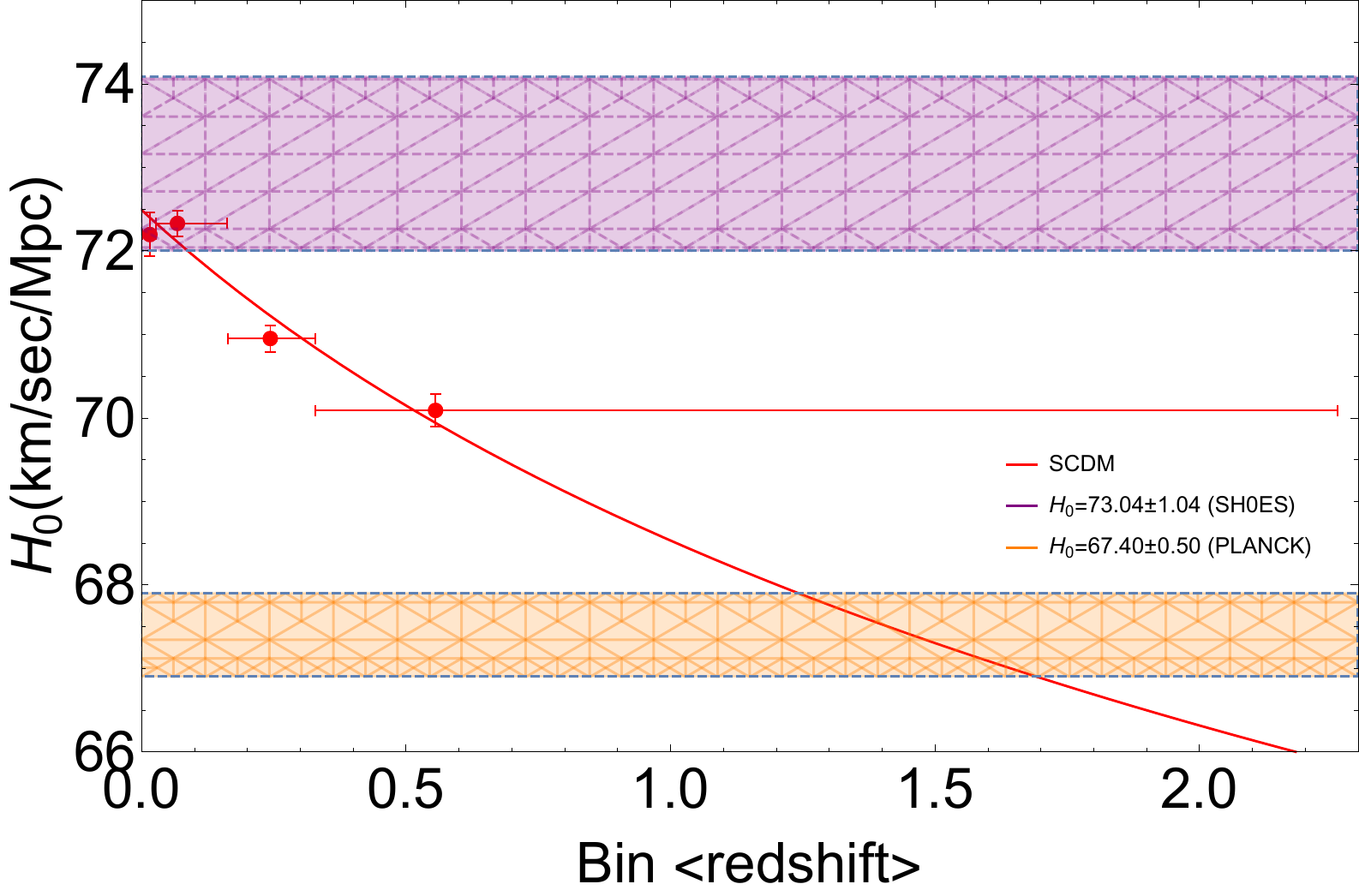}
    \includegraphics[width=0.75\linewidth]{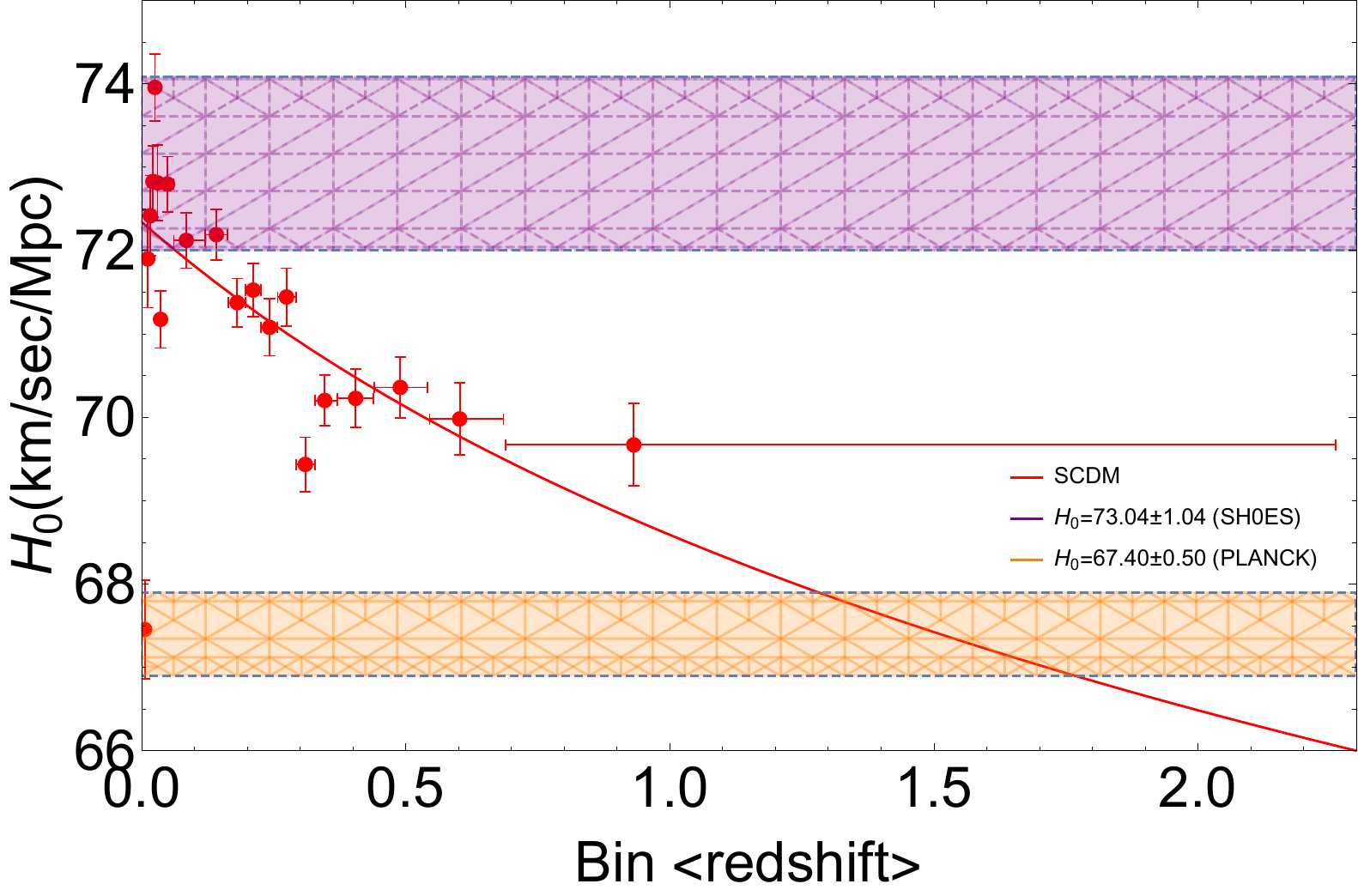}
    \includegraphics[width=0.75\linewidth]{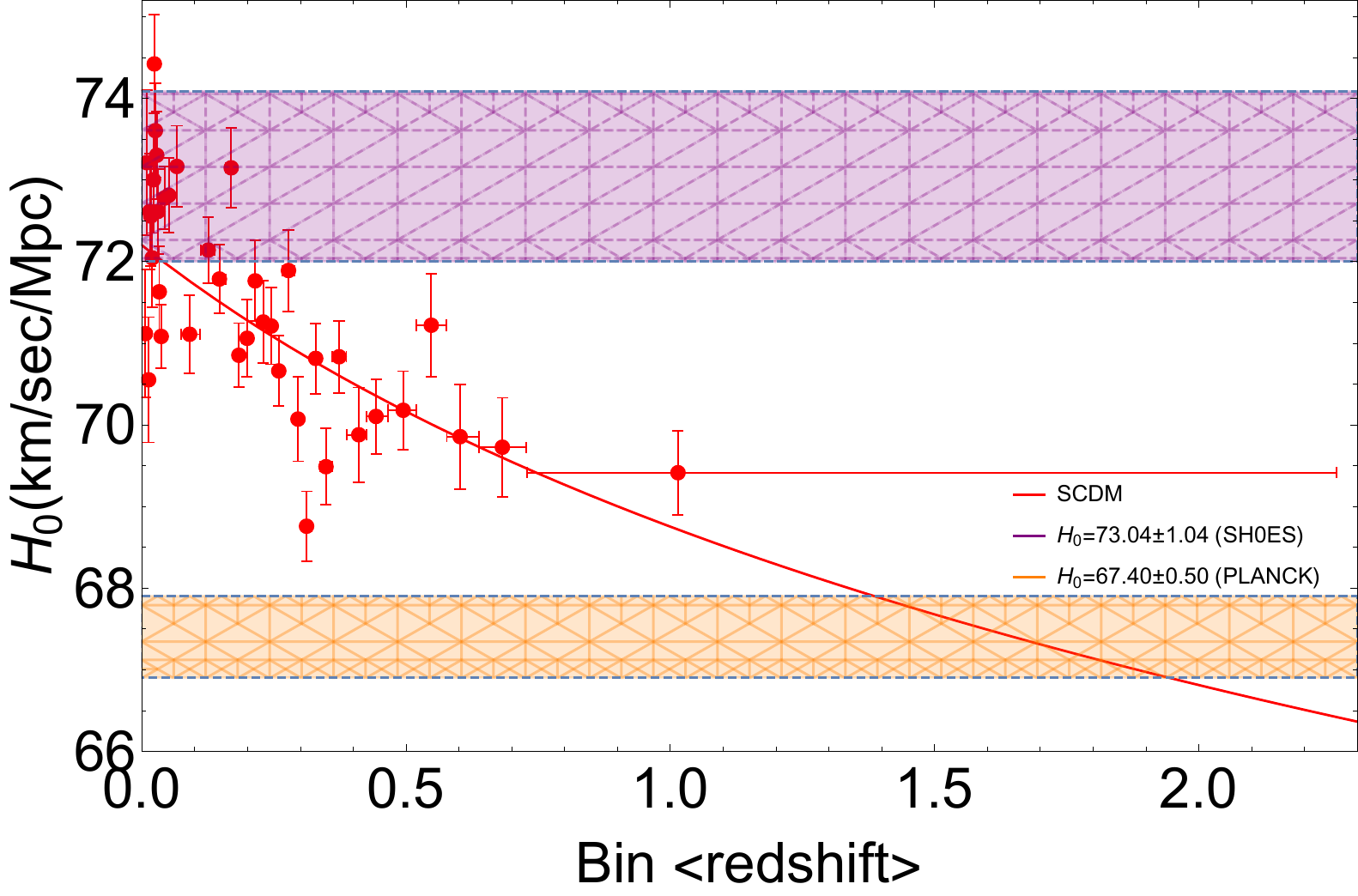}
    \caption{Fitting of $f(z)$ in the 3 bins (first panel), 4 bins (second panel), 20 bins (third panel) and 40 bins (fourth panel) of the Pantheon+ sample assuming the $S$CDM model. The purple and orange shaded areas have the same meaning as the ones in Figures \ref{fig:H0fit3binsP}, \ref{fig:H0fit4binsP}, \ref{fig:H0fit20binsP}, and \ref{fig:H0fit40binsP}.}
    \label{fig:H0fitallbinsPP}
\end{figure}

\begin{table}
\begin{centering}
\begin{tabular}{|c|c|c|c|c|}
\hline
\multicolumn{5}{|c|}{{\bf Pantheon+, $S$CDM, Varying $H_0$}}\tabularnewline
\hline
Bins & $H'_0$ & $\alpha$ & $\alpha/\sigma_{\alpha}$ & $\Delta_\%$ \\
& & & &
\tabularnewline
\hline
3 & $72.507\pm0.293$ & $0.078\pm0.016$ & $4.8$ & $7.0$ \tabularnewline
\hline
4 & $72.490\pm0.190$ & $0.081\pm0.011$ & $7.4$ & $6.5$  \tabularnewline
\hline
20 & $72.340\pm0.136$ & $0.077\pm0.008$ & $9.7$ & $2.9$ \tabularnewline
\hline
40 & $72.200\pm0.126$ & $0.071\pm0.007$ & $9.6$ & $1.8$ \tabularnewline
\hline
\end{tabular}
\caption{Fit parameters for $f(z)$ in the Pantheon+ bins, assuming the $S$CDM. First column: the number of Pantheon+ bins. Second column: the value of $H'_0 \pm \sigma_{H'_0}$ from the fitting. Third and fourth columns: the value of $\alpha \pm \sigma_{\alpha}$ and the number of $\sigma$ in which it is compatible with zero, respectively. Fifth column: the percentage difference.}
\label{tab:H0fitsPantheon+}
\par\end{centering}
\end{table}

\section{Summary and conclusions}\label{sec:conclusions}

A pronounced downward trend is detected in the $H_{0,k}$ estimates with mean redshift $z_k$ of binned Pantheon data. This trend appears reliable across different number of bins. 
This observed evolution of $H_0$ in the Pantheon sample is thereby confirmed consistent with the robustness in the analysis of $\Lambda$CDM \citep{Dainotti2021hubble,Dainotti2022hubble}. It is relevant to observe how extending this analysis to the Pantheon+ sample show very similar trends.
This trend is manifest in $\Delta_\%$ consistent with what is anticipated based on Fig. \ref{fig:H(z)/H0} when $K$ is modest, though somewhat less when $K$ is large. 
}

In light of the $H_0$-tension, the current findings (Tables \ref{tab:H0fits} and \ref{tab:H0fitsPantheon+}) in light of (\ref{EQN_q0b}) suggest searching for models with $q_0$ below that of $\Lambda$CDM, rather than $q_0$ closer to zero as in $S$CDM. Conversely, these results support the need for high-resolution observations of $q_0$. Though notoriously challenging to limit scatter in model-independent analysis of the LDL, these independent $q_0$ estimates \citep{cam2020a,cam2020b} are consistent with the present conclusions.

\section*{Acknowledgements}
BDS acknowledges the support for the accomodation from the National Astronomical Observatory of Japan (NAOJ). MvP acknowledges
NRF grant No. RS-2024- 00334550 of Korea.
BDS acknowledges the financial support from the University of Salerno, INFN - Gruppo Collegato di Salerno, and the FARB funding.

\bibliographystyle{elsarticle-harv} 
\bibliography{bibliography}

\end{document}